\documentclass[showpacs,preprintnumbers,prd,nofootinbib,floats,amssymb,floatfix]{revtex4-2}
\usepackage{graphicx}
\usepackage{amsxtra}
\usepackage{hyperref}
\usepackage{amssymb}
\usepackage{amstext}
\usepackage{amsmath}
\usepackage{cleveref}
\usepackage{stackrel}
\usepackage{blkarray}
\begin{document}
\title{ Hamiltonian analysis for  new  massive gravity }
\author{Alberto Escalante}  \email{aescalan@ifuap.buap.mx}
\author{P. Fernando Ocaña-Garc{\'i}a}  \email{pfgarcia@ifuap.buap.mx}
 \affiliation{Instituto de F\'isica, Benem\'erita Universidad Aut\'onoma de Puebla. \\ Apartado Postal J-48 72570, Puebla Pue., M\'exico, }

\begin{abstract}
A detailed  canonical analysis for  three-dimensional  massive
gravity  is performed.  The construction of the fundamental Dirac brackets,  the complete structure of the constraints and the counting of the physical degrees of freedom are reported. In addition, it is shown  that the extended Hamiltonian is  healed from   Orstrogradki's instabilities.     
\end{abstract}
 \date{\today}
\preprint{}
\maketitle

\section{Introduction}
It is well-known that topological massive gravity $[TMG]$ is a toy model, however,  it is a great laboratory for testing new ideas which  could be useful for the understanding of either the classical or quantum structure of gravity.  $TMG$  is constructed by the  Einstein-Hilbert  $[EH]$ action coupled with a  Chern-Simons term;  at the linearized regime it describes the propagation of a single massive state of helicity $\pm 2$ on a Minkowski background \cite{Deser, 2, 3}. The theory breaks parity and it is not-unitary. In fact,  it is claimed that if Newton's constant is negative,  then black hole states lead to non-unitary  theory. Moreover,  if Newton's constant is positive and the Chern-Simons coupling is not tuned to the chiral  point,  then massive excitation  lead also to non-unitarity. However, $TMG$ theory is a  dynamical model that is naively  power counting renormalizable; all these features make the theory so interesting \cite{3a, 3b, 3c, 4c}. \\
On the other hand, there are alternative models for describing the propagation of physical degrees of freedom in three dimensions, the so-called Minimal Massive Gravity $[MMG]$ and  New Massive Gravity $[NMG]$ \cite{ber2, 4, NMG}. $MMG$ is an extension of  $TMG$ that includes curvature-squared symmetric tensors and describes the propagation of a single local degree of freedom about an anti-de Sitter (AdS) vacuum, this propagation mode is physical; it is neither a tachyon nor a ghost. For this reason, $MMG$ is an attractive model due to it can be explored in the  (AdS) and quantum field  holographic correspondence. On the other hand,  $NMG$  is a parity-conserving  theory  and describes the propagation of two massive degrees of freedom of helicity $\pm 2$. These properties make  the theory very interesting  because it shares the same number of degrees of freedom with general relativity, considering  the clear exception that in three dimensions the physical degrees of freedom are massive. The theory is constructed by employing the $EH$ action together with  squared terms of the Ricci tensor and the Ricci scalar; due to the presence of these squared terms, it is a higher-order theory that could suffer from  Ostrogradski's instabilities. In this respect, the theory was analyzed at   Lagrangian level in \cite{4}, however,  a detailed Hamiltonian analysis has not been reported in the literature. A Hamiltonian description of the theory would allow us to understand how to exorcise the apparently  ghosts in $NMG$, and the canonical structure of the theory would help to make any progress in  the quantization program. The usual method used  for analyzing   higher-order theories is by employing the Ostrogradski-Dirac $[OD]$ method \cite{5, OD, OD1, 6, 7}. This is based on the extension of the phase space, where the fields and their temporal derivatives become the configuration variables, thus, it is introduced  a generalization of the canonical momenta for the higher-temporal derivatives of the fields. However, it is claimed that the $OD$  method does not allow direct identification of the complete structure of the constraints, so the constraints are fixed by hand to achieve consistency \cite{3}. \\
With all  commented above, in this paper we will perform the canonical analysis of $NMG$ by following a different scheme from that established in the $OD$ framework. To this end, a set of variables will be introduced to reduce the second-order time derivatives to  first-order. Then, by using the null vectors of the theory, the correct structure of the  constraints is  obtained without the need to fix them by hand as in  other approaches is done \cite{8, 9, 10}. Furthermore,  we  will  introduce  the Dirac brackets, thus the  second-class constraints will be useful  for either to build the extended Hamiltonian or   healing  the Ostrogradski sickness \cite{Pais}. It is worth commenting,  that usually  the extended Hamiltonian is  not built on a standard canonical analysis. In fact, if  there are non-trivial second-class constraints, then its construction is a difficult task. For  constructing  the extended Hamiltonian, we need to calculate the Dirac brackets and identify  all  Lagrange multipliers associated with the second-class constraints, then, the second class constraints are used for obtaining  the new structure of the Hamiltonian.  Thus, the structure of the extended Hamiltonian  will allow us to observe if  the Ostrogradski instability could  be healed. To complete the  analysis, we will study a close model  to $NMG$. This model will present a ghost and the  trivial structure of the second-class constraints will implies that the  Ostrogradski instability will not be healed, so the extended Hamiltonian will be unstable. \\
The paper is organized as follows: in Section II,  through the implementation of a set of auxiliary variables and their respective Lagrange multipliers, the canonical analysis of $NMG$ will be performed. The  correct structure of the constraints will be reported; with all constraints  at hand,  the Dirac brackets will be calculated  and the  extended Hamiltonian will be constructed. In addition, the counting of the physical degrees of freedom will be done and the   Dirac algebra between the extended Hamiltonian and the first-class constraints is reported. To highlight the role of the extended Hamiltonian in the identification of ghosts,  a close model to $NMG$  in Appendix A is studied. The same auxiliary variables of Section II are introduced and the extended Hamiltonian is also constructed, then the differences between the theories  are discussed. We finish the paper with the conclusions.\\




\section{Hamiltonian analysis of new massive gravity} 
We start our analysis with the following action   \cite{4, NMG}
\begin{equation}
 \label{eqn:ac}
 S[g_{\mu\nu}]=\frac{1}{\kappa^{2}}\int d^{3}x\hspace{0.1cm}\sqrt{-g}\left(R+\frac{1}{m^{2}}J\right), 
\end{equation}
where $g_{\mu\nu}$ is the metric tensor, $\alpha, \beta,...=0, 1, 2$; $\kappa$ is a constant with  dimension of mass  in fundamental units $[\kappa]=-1/2$, $m$ is a "relative" mass parameter and $J$ is a higher-order term  given by 
\begin{equation} \label{eqn:hi}
    J=R_{\mu\nu}R^{\mu\nu}-\frac{3}{8}R^{2}.
\end{equation}
 From the action, the following equations of motion arise 
\begin{equation} \label{eqn:em}
    J_{\mu\nu}+2m^{2}G_{\mu\nu}=0,
\end{equation}
where  $G_{\mu\nu}=R_{\mu\nu}-\frac{1}{2}Rg_{\mu\nu}$ is the Einstein tensor and $J_{\mu \nu}$ is given by 

\begin{equation}
     J_{\mu\nu}=2\Box R_{\mu\nu}-\frac{1}{2}\left(\nabla_{\nu}\nabla_{\mu}R+g_{\mu\nu}\Box R\right)-8R_{\mu}^{\rho}R_{\nu\rho}+\frac{9}{2}RR_{\mu\nu}+g_{\mu\nu}\left(3R_{\rho\sigma}R^{\rho\sigma}-\frac{13}{8}R^{2}\right),
\end{equation}
here $\Box$ is the D'Alambertian operator. We observe that the trace of  (\ref{eqn:em}) implies  
\begin{equation}
\label{eq5}
J= m^2R.
\end{equation}
For our aims, we will work by considering the standard  perturbation  of the metric around a Minkowski background, $ g_{\mu\nu}=\eta_{\mu\nu}+h_{\mu\nu}$. Hence, the  linearized versions of the Ricci tensor  and scalar curvature takes the form 
\begin{eqnarray}
\label{lin}
   \stackbin[Lin]{}{R_{\mu\nu}}&=&\frac{1}{2}\left(\partial_{\alpha}\partial_{\mu}h_{\nu}^{\alpha}+\partial_{\alpha}\partial_{\nu}h_{\mu}^{\alpha}-\partial_{\mu}\partial_{\nu}h-\partial_{\alpha} \partial^{\alpha}h_{\mu\nu}\right), \nonumber \\  
      \stackbin[Lin]{}R&=&\partial_{\mu}\partial_{\nu}h^{\mu\nu}-\partial_{\alpha} \partial^{\alpha}h.
\end{eqnarray}
In this manner, by using (\ref{lin}) into   (\ref{eqn:em}) and (\ref{eq5}),  the following linearized equations of motion arise 
\begin{eqnarray}\label{eqn:lin}
   (\Box +m^2)\stackbin[Lin]{}{G_{\mu\nu}} &=&0, \nonumber \\
   \stackbin[Lin]{}{R}&=&0.
\end{eqnarray}
 We observe that the first equation of motion is a Klein-Gordon equation for the linerarized Einstein tensor. The second one,  is an implicit constraint  that at Hamiltonian level enforces the presence of second-class constraints, this  will be shown below. \\
 Furthermore,  another way for obtaining the linearized equations (\ref{eqn:lin})  is by the variation of the linearized  version of (\ref{eqn:ac}), which gives the following  Lagrangian 
\begin{eqnarray}\label{Llin}
    \nonumber 
       \mathcal{L}&=&\frac{1}{2}\left(\partial_{\mu}h^{\mu\nu}\partial_{\alpha}h_{\nu}^{\alpha}-\frac{1}{2}\partial^{\alpha} h^{\mu\nu}\partial_{\alpha}h_{\mu\nu} -\partial_{\nu}h\partial_{\mu}h^{\mu\nu}+\frac{1}{2}\partial_{\alpha}h \partial^{\alpha}h\right)+\frac{1}{4m^{2}}\Big(\frac{1}{2}\partial_{\mu}\partial_{\nu}h^{\mu\nu}\partial_{\alpha}\partial_{\beta}h^{\alpha\beta} \nonumber\\ 
       &+& \Box h_{\mu\nu}\Box h^{\mu\nu}-\frac{1}{2}\Box h\Box h +\partial_{\mu}\partial_{\nu}h^{\mu\nu}\Box h-2\partial_{\alpha}\partial_{\mu}h_{\nu}^{\alpha}\Box h^{\mu\nu}\Big).
\end{eqnarray}
The Lagrangian (\ref{Llin}) will be our subject of study. In fact, by performing the $2+1$ decomposition, we obtain the following form of the Lagrangian
\begin{eqnarray}\label{l11}
\nonumber
        \mathcal{L}&=&\frac{1}{4}\Dot{h}_{ij}\Dot{h}^{ij}-\frac{1}{4}\dot{h}_{i}^{i}\dot{h}_{j}^{j}-\Dot{h}_{ij}\partial^{i}h_{0}^{j}-\dot{h}_{i}^{i}\partial_{k}h^{0k}+\frac{1}{4m^{2}}\Big( \ddot{h}^{ij}\ddot{h}_{ij}-2\ddot{h}_{ij}\nabla^{2} h^{ij}-\frac{1}{2}\ddot{h}_{i}^{i}\ddot{h}_{j}^{j}+\ddot{h}_{i}^{i}\nabla^{2} h_{j}^{j} \nonumber \\
        &-&4\ddot{h}_{ij}\partial^{j}\Dot{h}_{0}^{i}+2\ddot{h}_{ij}\partial_{l}\partial^{j}h^{il}+4\partial^{j}\partial^{k} h_{0}^{i}\partial_{k}\Dot{h}_{ij}-2\partial_{j}\Dot{h}_{0i}\partial^{i}\partial_{l}h^{jl}-2\partial_{j}\Dot{h}_{0i}\partial^{i}\partial^{j} h_{k}^{k}-\ddot{h}_{00}\nabla^{2} h_{i}^{i}\nonumber \\
        &-&2\partial_{k}\Dot{h}^{0k}\ddot{h}_{i}^{i}-\partial_{l}\partial_{m}h^{lm}\ddot{h}_{i}^{i}-2\ddot{h}_{0i}\nabla^{2} h^{0i}-2\partial^{i}\partial^{j} h_{00}\partial_{j}\Dot{h}_{0i}+2\partial_{i}\partial_{j}h^{ij}\ddot{h}_{00}\Big)-V,
\end{eqnarray}
where 
\begin{equation}\label{l2}
    \begin{split}
    \nonumber
        V&=\frac{1}{2}\partial_{i}h_{0j}\partial^{i}h^{0j}-\frac{1}{2}\partial^{i}h^{j0}\partial_{j}h_{i0}-\frac{1}{2}\partial_{i}h_{00}\partial_{j}h^{ij}+\frac{1}{2}\partial_{i}h_{k}^{k}\partial_{j}h^{ij}+\frac{1}{2}\partial_{i}h_{00}\partial^{i}h_{k}^{k}-\frac{1}{4}\partial_{i}h_{j}^{j}\partial^{i}h_{k}^{k}\\&\hspace{4mm}-\frac{1}{2}\partial^{i}h^{jk}\partial_{j}h_{ik}+\frac{1}{4}\partial_{i}h_{jk}\partial^{i}h^{jk}-\frac{1}{4m^{2}}\left(\partial_{i}\partial_{j}h^{ij}\nabla^{2} h_{k}^{k}+2\nabla^{2} h^{0i}\nabla^{2} h_{0i}-2\nabla^{2} h^{ij}\partial_{k}\partial_{i}h^{k}_{j}\right.\\&\left.\hspace{4mm}+\frac{1}{2}\left(\nabla^{2} h^{00}\right)^{2}+\nabla^{2} h_{00}\nabla^{2} h_{i}^{i}-\partial_{i}\partial_{j}h^{ij}\nabla^{2} h_{00}+\nabla^{2} h^{ij}\nabla^{2} h_{ij}-\frac{1}{2}(\nabla^{2} h_{i}^{i})^{2}+\frac{1}{2}(\partial_{i}\partial_{j}h^{ij})^{2} \right.\\&\left.\hspace{4mm}-2\nabla^{2} h^{j0}\partial_{i}\partial_{j}h^{i}_{0}\right),
    \end{split}
\end{equation}
here $\nabla^{2}=\partial_{i}\partial^{i}$. As far as we know, a complete  canonical analysis of the Lagrangian  (\ref{l11}) has not been reported in the literature. In this respect,  we have commented above that the standard way  for analyzing  a higher-order theory is by using the   $OD$  framework \cite{OD, OD1}, however, this approach is  long and the constraints are not under control. So, we will perform our analysis by introducing the following  variables \cite{11, 12}
\begin{equation}\label{extrinsic}
    K_{ij}=\frac{1}{2}\left(\dot{h}_{ij}-\partial_{i}h_{0j}-\partial
    _{j}h_{0i}\right),
\end{equation}
these variables are an extrinsic curvature type. In this way,  the  Lagrangian is rewritten in the following new fashion
\begin{eqnarray}\label{ll1}
\nonumber 
 \mathcal{L}&=&K_{ij}K^{ij}-K^{2}-\frac{1}{2}h^{00}R_{ij}^{\hspace{2.5mm}ij}-\frac{1}{2}h^{ij}\left(R_{ikj}^{\hspace{3mm}k}-\frac{1}{2}\delta_{ij}R_{lm}^{\hspace{2.5mm}lm}\right)+\frac{1}{m^{2}}\left(\dot{K}^{ij}\dot{K}_{ij}-\frac{1}{2}\dot{K}^{2}\right.\\ \nonumber 
        &&-\frac{3}{2}\Dot{K}R_{ij}^{\hspace{2mm}ij}-\frac{1}{2}\dot{K}\nabla^{2} h_{00}+\dot{K}^{ij}\partial_{i}\partial_{j}h_{00}+2\dot{K}_{ij}R_{\hspace{1mm}l}^{i\hspace{2mm}jl}-2\partial^{l}K_{il}\partial^{j}K^{i}_{j}+4\partial^{j}K_{ij}\partial^{i}K\\ \nonumber 
        & &\left.-2\partial_{i}K\partial^{i}K+R_{ikj}^{\hspace{4mm}k}R_{\hspace{1mm}l}^{i\hspace{2mm}jl}+\partial_{i}\partial_{j}h_{00}R_{\hspace{1mm}k}^{i\hspace{2mm}jk}-\frac{3}{8}R_{ij}^{\hspace{2mm}ij}R_{lm}^{\hspace{3mm}lm}-\frac{3}{4}\nabla^{2} h_{00}R_{ij}^{\hspace{2mm}ij}+\frac{1}{8}(\nabla^{2} h_{00})^{2}\right)\\ 
        && +\alpha^{ij}\left(\dot{h}_{ij}-2\partial_{i}h_{0j}-2K_{ij}\right),
 \end{eqnarray}
where $\alpha^{ij}$ are  Lagrange multipliers enforcing the relation (\ref{extrinsic}) and
\begin{equation}
 \nonumber R_{ijkl}=-\frac{1}{2}\left[ \partial_i\partial_k h_{jl}+\partial_j \partial_lh_{ik}-\partial_i \partial_l h_{jk}-\partial_j \partial_k h_{il} \right].
 \end{equation}
The new canonical variables  of the system are  given by $h_{\mu\nu}$, $K_{ij}$, $\alpha_{ij}$ and their  corresponding canonical momenta given by $\pi^{\mu\nu}$, $P^{ij}$ and $\tau^{ij}$. From the definition of the momenta, we identify the following relations 
\begin{eqnarray}\label{eqcan}
\nonumber 
        \pi^{00}&=&\frac{\partial \mathcal{L}}{\partial \dot{h}_{00}}=0,\\ \nonumber 
        \pi^{0i}&=&\frac{\partial \mathcal{L}}{\partial \dot{h}_{0i}}=0,\\ \nonumber 
        \pi^{ij}&=&\frac{\partial \mathcal{L}}{\partial \dot{h}_{ij}}=\alpha^{ij},\\ \nonumber 
        \tau^{ij}&=&\frac{\partial\mathcal{L}}{\partial \dot{\alpha}_{ij}}=0,\\ 
        P^{ij}&=&\frac{\partial \mathcal{L}}{\partial \dot{K}_{ij}}=\frac{1}{m^{2}}\left(-\delta^{ij}\dot{K}+2\dot{K}^{ij}-\frac{1}{2}\delta^{ij}\nabla^{2} h_{00}+\partial^{i}\partial^{j}h_{00}+2R_{\hspace{1mm}l}^{i\hspace{2mm}jl}-\frac{3}{2}\delta^{ij}R_{lm}^{\hspace{2mm}lm}\right),
\end{eqnarray}
and  the fundamental Poisson brackets between the canonical variables will be  
\begin{eqnarray}
\nonumber
    \left\lbrace h_{\mu\nu},\pi^{\alpha\beta}\right\rbrace&=&\frac{1}{2}\left(\delta_{\mu}^{\alpha}\delta_{\nu}^{\beta}+\delta_{\mu}^{\beta}\delta_{\nu}^{\alpha}\right)\delta^{2}(x-y), \nonumber \\
    \left\lbrace K_{ij},P^{lm}\right\rbrace&=&\frac{1}{2}\left(\delta_{i}^{l}\delta_{j}^{m}+\delta_{i}^{m}\delta_{j}^{l}\right)\delta^{2}(x-y), \nonumber \\ 
     \left\lbrace\alpha_{ij},\tau^{lm}\right\rbrace&=&\frac{1}{2}\left(\delta_{i}^{l}\delta_{j}^{m}+\delta_{i}^{m}\delta_{j}^{l}\right)\delta^{2}(x-y).
\end{eqnarray}
From  the Lagrangian (\ref{ll1}) and the canonical momenta (\ref{eqcan}), we can construct the canonical  Hamiltonian, it is given by 
\begin{equation}
    \begin{split}
      \mathcal{H}_c&=\tau^{ij}\dot{\alpha}_{ij}+\pi^{ij}\dot{h}_{ij}+P^{ij}\dot{K}_{ij}-\mathcal{L}\\
      &=\frac{m^{2}}{4}P^{ij}P_{ij}+\frac{1}{4}P\nabla^{2} h_{00}-\frac{1}{2}P^{ij}\partial_{i}\partial_{j}h_{00}-P^{ij}R_{ilj}^{\hspace{3mm}l}+\frac{3}{4}PR_{lm}^{\hspace{3mm}lm}+
       K^{2}-K_{ij}K^{ij}+2\pi^{ij}K_{ij}\\ \nonumber 
       &\hspace{4mm}+\frac{1}{2}h_{00}R_{ij}^{\hspace{2.5mm}ij}+\frac{1}{2}h^{ij}\left(R_{ikj}^{\hspace{3mm}k}-\frac{1}{2}\delta_{ij}R_{lm}^{\hspace{2.5mm}lm}\right)+\frac{1}{m^{2}}\left(2\partial^{l}K_{il}\partial^{j}K^{i}_{j}-4\partial^{j}K_{ij}\partial^{i}K+2\partial_{i}K\partial^{i}K\right.\\  
       &\left.\hspace{4mm}+\frac{1}{4}\nabla^{2} h_{00}R_{ij}^{\hspace{2mm}ij}\right)-2\partial_i\pi^{ij}h_{0j},
    \end{split}
\end{equation}
we can observe the presence of linear terms in the conjugate momenta $\pi^{ij}$, which could be related to  an Ostrogradski's instability. However, in further lines we will remove this apparently instability by means the Dirac brackets and the second-class constraints. \\
On the other hand, from the definition of the momenta, we identify the following primary constraints 
\begin{eqnarray}
\label{conpri}
  \nonumber 
      \phi^{00}&:&\pi^{00}\approx0, \nonumber \\
      \phi^{0i}&: &\pi^{0i}\approx0, \nonumber \\
      \phi^{ij}&:&\pi^{ij}-\alpha^{ij}\approx0, \nonumber \\
      \psi^{ij}&:&\tau^{ij}\approx0, \nonumber \\
      \psi&:&P+\frac{1}{m^{2}}R_{lm}^{\hspace{2mm}lm}\approx0,
\end{eqnarray}
we observe from the definition of the momenta (\ref{eqcan}) that the velocity $\dot{K}$ can not be expressed in terms of the canonical momenta, then the constraint $\psi$  arise. The constraint $\psi$ is not trivial and  it will be  of second-class, this constraint will   be useful at the end of the analysis for removing the apparent Ostrogradski's instability. On the other hand, we calculate  the non-zero Poisson brackets between primary constraints which are given by  
\begin{eqnarray}
\nonumber 
   \left\lbrace \psi^{ij},\phi^{lm}\right\rbrace &=&\frac{1}{2}\left(\delta^{il}\delta^{jm}+\delta^{im}\delta^{jl}\right)\delta^{2}(x-y), \nonumber \\
   \left\lbrace \psi,\phi^{ij}\right\rbrace &=&\frac{1}{m^{2}}\left(\partial^{i}\partial^{j}-\delta^{ij}\nabla^{2}\right)\delta^{2}(x-y).
\end{eqnarray}
Thus, the  primary Hamiltonian reads 
\begin{eqnarray}\label{primaryHamiltonian}
    \nonumber 
      \mathcal{H}'&=&\frac{m^{2}}{4}P^{ij}P_{ij}-\frac{1}{2}P^{ij}\partial_{i}\partial_{j}h_{00}-P^{ij}R_{ilj}^{\hspace{3mm}l}+\frac{3}{4}PR_{lm}^{\hspace{3mm}lm}+\frac{2}{m^{2}}\partial^{l}K_{il}\partial^{j}K^{i}_{j}-\frac{4}{m^{2}}\partial^{j}K_{ij}\partial^{i}K \nonumber \\
      &&+\frac{2}{m^{2}}\partial_{i}K\partial^{i}K+
       K^{2}-K_{ij}K^{ij}+\frac{1}{2}h^{ij}\left(R_{ikj}^{\hspace{3mm}k}-\frac{1}{2}\delta_{ij}R_{lm}^{\hspace{2.5mm}lm}\right)+\frac{1}{2}h_{00}R_{ij}^{\hspace{2.5mm}ij}+2\pi^{ij}K_{ij}\nonumber \\
       &&+2\pi^{ij}\partial_{i}h_{0j}+u_{\mu\nu}\phi^{\mu\nu}+\zeta_{ij}\psi^{ij}+\xi\psi,
\end{eqnarray}
where $u_{\mu\nu}$, $\zeta_{ij}$ and $\xi$ are Lagrange multipliers enforcing the primary constraints. With the primary Hamiltonian,  we calculate the consistency conditions  on the primary constraints, thus,  we obtain 
\begin{eqnarray}
\nonumber 
    S&:& \dot{\phi}^{00}=\frac{1}{2}\left(\partial_{i}\partial_{j}P^{ij}- R_{ij}^{\hspace{2mm}ij}\right)\approx 0,\nonumber \\
    S^{i}&:&\dot{\phi}^{0i}=\partial_{j}\pi^{ij}\approx 0, \nonumber\\
    S^{lm}&:&\dot{\phi}^{lm}=\frac{1}{2}\left(\partial^{l}\partial_{i}P^{im}+\partial^{m}\partial_{i}P^{il}-\nabla^{2}P^{lm}-\delta^{lm}\partial_{i}\partial_{j}P^{ij}\right)-\left(\partial^{l}\partial^{m}-\delta^{lm}\nabla^{2}\right)\left(\frac{3}{4}P+\frac{1}{2}h_{00}\right), \nonumber\\ 
    &-&\left(R_{\hspace{1mm}k}^{l\hspace{1.5mm}mk}-\frac{1}{2}\delta^{lm}R_{ij}^{\hspace{2mm}ij}\right)-\frac{1}{m^{2}}\left(\partial^{l}\partial^{m}-\delta^{lm}\nabla^{2}\right)\xi-\zeta^{lm}\approx0, \nonumber\\
    Q^{ij}&:&\dot{\psi}^{ij}=u^{ij}\approx0, \nonumber \\
    W&:& \dot{\psi}=\frac{2}{m^{2}}\left(\nabla^{2}K-\partial_{i}\partial_{j}K^{ij}\right)-2K-2\pi_{i}^{i}\approx0,
\end{eqnarray}
from these equations we can observe  that $S, S^i, W$ are secondary constraints while  $S^{ij}, Q^{ij}$ only give  relations between the Lagrange multipliers. From   consistency of the secondary constraints we find that 
\begin{eqnarray}\label{ter}
\nonumber
     \dot{S}&=&-\partial_{i}\partial_{j}\pi^{ij}\approx 0, \nonumber\\
     \dot{S}^{i}&\approx& 0, \nonumber\\
     \dot{W}&=&-\partial_{i}\partial_{j}P^{ij}-4\xi\approx 0,
\end{eqnarray}
 we can observe that  not further  constraints arise. Hence,  the complete set of constraints is given by 
\begin{eqnarray}
\nonumber 
      \phi^{00}&:&\pi^{00}\approx 0, \nonumber \\
      \phi^{0i}&:& \pi^{0i}\approx 0, \nonumber \\
      \phi^{ij}&:&\pi^{ij}-\alpha^{ij}\approx0, \nonumber \\
      \psi^{ij}&:&\tau^{ij}\approx 0, 	\nonumber \\
      \psi&:&P+\frac{1}{m^{2}}R_{lm}^{\hspace{2mm}lm}  \approx0, \nonumber\\
      S&:&\partial_{i}\partial_{j}P^{ij}- R_{ij}^{\hspace{2mm}ij} \approx 0, \nonumber \\
      S^{i}&:&\partial_{j}\pi^{ij}\approx 0, \nonumber \\
      W&:&\frac{1}{m^{2}}\left(\partial_{i}\partial_{j}K^{ij}-\nabla^{2}K\right)+K+\pi_{i}^{i} \approx 0. 
\end{eqnarray}
With all constraints at hand, we can perform their  classification into first and second-class. For this step we calculate the following matrix,  whose entries are the Poisson brackets between all constraints  
\begin{equation}\label{matrix2}
\makeatletter\setlength\BA@colsep{7pt}\makeatother
\arraycolsep=1.4pt\def\arraystretch{0.85}
   M=
    \begin{blockarray}{ccccccccc}
        & \phi^{00} & \phi^{0i}  & \phi^{ij}  & \psi^{ij} & \psi & S & S^{i} & W \\
      \begin{block}{c(cccccccc)}
        \phi^{00} & 0 & 0 & 0 & 0 & 0 & 0 & 0 & 0 \\
        \phi^{0l}  & 0 & 0 & 0 & 0 & 0 & 0 & 0 & 0\\
        \phi^{lm}  & 0 & 0 & 0 & \left\lbrace \phi^{lm},\psi^{ij}\right\rbrace  & \left\lbrace \phi^{lm},\psi\right\rbrace & \left\lbrace \phi^{lm},S\right\rbrace & 0 & 0   \\
        \psi^{lm}   & 0 & 0 & \left\lbrace \psi^{lm},\phi^{ij}\right\rbrace  & 0 & 0 & 0 & 0 & 0 \\
        \psi  & 0 & 0 & \left\lbrace \psi,\phi^{ij}\right\rbrace & 0 & 0 & 0 & 0 & \left\lbrace \psi,W\right\rbrace \\ 
        S & 0 & 0 &\left\lbrace S,\phi^{ij}\right\rbrace & 0 & 0 & 0 & 0 & 0 \\
        S^{l} & 0 & 0 & 0  & 0 & 0 & 0 & 0 & 0  \\
        W  & 0 & 0 & 0 & 0 & \left\lbrace W,\psi\right\rbrace & 0 & 0 & 0 \\
      \end{block}
    \end{blockarray}
\end{equation}
where the non-zero Poisson brackets between the constraints are given by 
\begin{eqnarray}
\nonumber
        \left\lbrace \psi^{ij},\phi^{lm}\right\rbrace &=&\frac{1}{2}\left(\delta^{il}\delta^{jm}+\delta^{im}\delta^{jl}\right)\delta^{2}(x-y), \nonumber\\
        \left\lbrace \psi,\phi^{ij}\right\rbrace &=&\frac{1}{m^{2}}\left(\partial^{i}\partial^{j}-\delta^{ij}\nabla^{2}\right)\delta^{2}(x-y), \nonumber\\
        \left\lbrace S,\phi^{ij}\right\rbrace&=&-\left(\partial^{i}\partial^{j}-\delta^{ij}\nabla^{2}\right)\delta^{2}(x-y), \nonumber\\
         \left\lbrace \psi, W\right\rbrace&=&-2\delta^{2}(x-y).
\end{eqnarray}
After a long calculation, we can see that the matrix (\ref{matrix2})  has a rank= 8 and  6 null vectors. From the null vectors we find the following 6 first-class constraints  
\begin{eqnarray}
    \nonumber
      \Gamma_{1}&:&\pi^{00}\approx0, \nonumber\\
      \Gamma_{2}^{i}&:&\pi^{0i}\approx0, \nonumber
      \\
      \Gamma_{3}^{i}&:&
     \partial_{j}\pi^{ij}\approx 0, \nonumber \\
    \Gamma_{4}&:&\partial_{i}\partial_{j}P^{ij}- R_{ij}^{\hspace{2mm}ij}+\left(\partial_{i}\partial_{j}-\delta_{ij}\nabla^{2}\right)\tau^{ij}\approx 0,
\end{eqnarray}
which allow us to  identify the following  8 second-class constraints 
\begin{eqnarray}\label{seconclass}
  \nonumber
      \chi_{1}^{ij}&:&\pi^{ij}-\alpha^{ij}\approx0, \nonumber\\
      \chi_{2}^{ij}&:&\tau^{ij}\approx0, \nonumber \\
       \chi_{3}&:&P+\frac{1}{m^{2}}R_{ij}^{\hspace{2mm}ij} \approx0, \nonumber \\
      \chi_{4}&: &K+\frac{1}{m^{2}}\left(\partial_{i}\partial_{j}K^{ij}-\nabla^{2}K\right)+\pi_{i}^{i}\approx 0.
\end{eqnarray}
where $\Gamma_{3}^{i}$ is the so-called Gauss constraint. With the classification of the constraints, we carry out the counting of physical degrees of freedom  as follows: there are $24$ canonical variables, eight  second-class constraints and six first-class constraints, thus 
\begin{equation}
    DOF=\frac{1}{2}\left(24-8-2*6\right)=2, 
\end{equation} 
 these degrees of freedom corresponds to massive modes of helicities $\pm2$ \cite{4}. \\
 We shall   introduce the  Dirac brackets. To achieve this, we first calculate  the algebra between the second-class constraints, it is given by 
\begin{eqnarray}
\nonumber
    \left\lbrace \chi_{1}^{ij},\chi_{2}^{lm}\right\rbrace&=&-\frac{1}{2}\left(\delta^{il}\delta^{jm}+\delta^{im}\delta^{jl}\right)\delta^{2}(x-y), \nonumber\\
    \left\lbrace \chi_{1}^{ij},\chi_{3}\right\rbrace&=&-\frac{1}{m^{2}}\left(\partial^{i}\partial^{j}-\delta^{ij}\nabla^{2}\right)\delta^{2}(x-y), \nonumber
    \\
    \left\lbrace \chi_{3},\chi_{4}\right\rbrace&=&-2\delta^{2}(x-y),
\end{eqnarray}
and we express it in matrix form as follows
\begin{equation}\label{matrix3}
\makeatletter\setlength\BA@colsep{7pt}\makeatother
\arraycolsep=1.4pt\def\arraystretch{0.85}
   C_{\alpha\beta}=
    \begin{blockarray}{ccccccccc}
        & \chi_{1}^{11} & \chi_{1}^{12}  & \chi_{1}^{22} & \chi_{2}^{11} & \chi_{2}^{12} & \chi_{2}^{22} & \chi_{3} &\chi_{4} \\
      \begin{block}{c(cccccccc)}
        \chi_{1}^{11} & 0 & 0   & 0 & -m^{2} & 0 & 0 & \partial^{2}\partial^{2}& 0  \\
        \chi_{1}^{12} & 0 & 0   & 0 & 0 & -\frac{1}{2}m^{2} & 0 & -\partial^{1}\partial^{2}  & 0\\
        \chi_{1}^{22} & 0 & 0   & 0 & 0 & 0 & -m^{2} & \partial^{1}\partial^{1}    & 0 &   \\
        \chi_{2}^{11} & m^{2} & 0   & 0 & 0 & 0 & 0 & 0 & 0 \\
        \chi_{2}^{12} & 0 & \frac{1}{2}m^{2}   & 0  & 0 & 0 & 0 & 0 & 0\\
        \chi_{2}^{22} & 0 & 0   & m^{2} & 0 & 0 & 0 & 0 & 0 \\ 
        \chi_{3} & -\partial^{2}\partial^{2} & \partial^{1}\partial^{2} & -\partial^{1}\partial^{1} & 0 & 0 & 0 & 0 &-2m^{2} \\
        \chi_{4} & 0 & 0 & 0 & 0 & 0  & 0 & 2m^{2} & 0  \\
      \end{block}
    \end{blockarray}\hspace{1mm}\frac{1}{m^{2}}\delta^{2}(x-y).
\end{equation}
It is well known that the Dirac brackets play an important role in the quantization of any theory with second-class constraints. In fact, they  are promoted to commutators and they can be used for the identification of  observables. On the other hand, at classical level either the Dirac brackets or the second class constraints can be  used for constructing the extended Hamiltonian.  It is worth commenting,  that the equations of motion obtained by means of the extended
Hamiltonian  are mathematically different from the Euler–Lagrange
equations, but the difference is unphysical, thus, the construction of the extended Hamiltonian is important. In this manner,  the Dirac  brackets are defined as

\begin{equation}
    \left\lbrace A, B\right\rbrace_{D}= \left\lbrace A, B\right\rbrace-\int dudv\left\lbrace A, \chi_{\alpha}(u)\right\rbrace C^{\alpha \beta}\left\lbrace \chi_{\beta}(v), B\right\rbrace,
\end{equation}
where $C^{\alpha\beta}$ is the inverse of (\ref{matrix3}) given by 

\begin{equation}\label{inverse}
\makeatletter\setlength\BA@colsep{7pt}\makeatother
\arraycolsep=1.4pt\def\arraystretch{0.85}
   C^{\alpha\beta}=
    \begin{blockarray}{ccccccccc}
        & \chi_{1}^{11} & \chi_{1}^{12}  & \chi_{1}^{22} & \chi_{2}^{11} & \chi_{2}^{12} & \chi_{2}^{22} & \chi_{3} &\chi_{4} \\
      \begin{block}{c(cccccccc)}
        \chi_{1}^{11} & 0 & 0   & 0 & m^{2} & 0 & 0 & 0 & 0  \\
        \chi_{1}^{12} & 0 & 0   & 0 & 0 & 2m^{2} & 0 & 0 & 0\\
        \chi_{1}^{22} & 0 & 0   & 0 & 0 & 0 & m^{2} & 0 & 0  \\
        \chi_{2}^{11} & -m^{2} & 0   & 0 & 0 & 0 & 0 & 0 & \frac{1}{2}\partial_{2}\partial_{2} \\
        \chi_{2}^{12} & 0 &-2m^{2}   & 0 & 0 & 0 & 0 & 0 & -\partial_{1}\partial_{2}\\
        \chi_{2}^{22} & 0 & 0   & -m^{2} & 0 & 0 & 0 & 0 & \frac{1}{2}\partial_{1}\partial_{1} \\ 
        \chi_{3} & 0 & 0 & 0 & 0 & 0 & 0 & 0 & \frac{1}{2}m^{2}\\
        \chi_{4} & 0 & 0   & 0 & -\frac{1}{2}\partial_{2}\partial_{2} & \partial_{1}\partial_{2}  & -\frac{1}{2}\partial_{1}\partial_{1} & - \frac{1}{2}m^{2} & 0  \\
      \end{block}
    \end{blockarray}\hspace{1mm}\frac{1}{m^{2}}\delta^{2}(x-y).
\end{equation}
Hence,  we obtain  the following  non-trivial Dirac brackets
\begin{eqnarray}
\nonumber
    \left\lbrace h_{ij},\pi^{lm}\right\rbrace_{D}&=&\frac{1}{2}\left(\delta_{i}^{l}\delta_{j}^{m}+\delta_{i}^{m}\delta_{j}^{l}\right)\delta^{2}(x-y)+\frac{1}{2m^{2}}\delta_{ij}\left(\partial^{l}\partial^{m}-\delta^{lm}\nabla^{2}\right)\delta^{2}(x-y), \nonumber \\
    \left\lbrace h_{ij},\alpha^{lm}\right\rbrace_{D}&=&\frac{1}{2}\left(\delta_{i}^{l}\delta_{j}^{m}+\delta_{i}^{m}\delta_{j}^{l}\right)\delta^{2}(x-y)+\frac{1}{2m^{2}}\delta_{ij}\left(\partial^{l}\partial^{m}-\delta^{lm}\nabla^{2}\right)\delta^{2}(x-y), \nonumber
   \\
     \left\lbrace K_{ij},P^{lm}\right\rbrace_{D}&=&\frac{1}{2}\left(\delta_{i}^{l}\delta_{j}^{m}+\delta_{i}^{m}\delta_{j}^{l}\right)\delta^{2}(x-y)-\frac{1}{2}\delta_{ij}\left(\delta^{lm}+\frac{1}{m^{2}}\left(\partial^{l}\partial^{m}-\delta^{lm}\nabla^{2}\right)\right)\delta^{2}(x-y), \nonumber \\
     \left\lbrace \pi_{ij},P^{lm}\right\rbrace_{D}&=&\frac{1}{2m^{2}}\left(\partial_{i}\partial_{j}-\delta_{ij}\nabla^{2}\right)\left(\delta^{lm}+\frac{1}{m^{2}}\left(\partial^{l}\partial^{m}-\delta^{lm}\nabla^{2}\right)\right)\delta^{2}(x-y), \nonumber\\
     \left\lbrace
     \alpha_{ij},P^{lm}\right\rbrace_{D}&=&\frac{1}{2m^{2}}\left(\partial_{i}\partial_{j}-\delta_{ij}\nabla^{2}\right)\left(\delta^{lm}+\frac{1}{m^{2}}\left(\partial^{l}\partial^{m}-\delta^{lm}\nabla^{2}\right)\right)\delta^{2}(x-y),\nonumber \\
      \left\lbrace
     h_{ij},K^{lm}\right\rbrace_{D}&=&-\frac{1}{2}\delta_{ij}\delta^{lm}\delta^{2}\left(x-y\right).
\end{eqnarray}
Now, we shall  construct  the extended Hamiltonian,  which is a fundamental element in the canonical formulation. In fact, the extended Hamiltonian is a first-class function and it is used in the quantization program  due to it  contains  all  relevant information of  the theory. The extended Hamiltonian is defined  by 
\begin{equation}
    H_{E}=H+w_{\alpha}\chi^{\alpha},
\end{equation}
where $w_{\alpha}$  are the  Lagrange multipliers associated with the second-class constraints, these multipliers can be determined through  \cite{13, 14}
\begin{equation}
    w_{\alpha}=C_{\beta\alpha}^{-1}\left\lbrace \chi^{\beta},H\right\rbrace_D.
\end{equation}
In this manner, by using the second-class constraints (\ref{seconclass}) and the matrix (\ref{inverse}) the following expressions for the Lagrange multipliers are obtained 
\begin{eqnarray}
\label{multiplier}
  \nonumber
        w_{1}&=&0, \nonumber \\
        w_{2}&=&0, \nonumber \\
        w_{3}&=&0, \nonumber \\
        w_{4}&=& \left(\partial^{1}\partial_{i}P^{i1}-\frac{1}{2}\nabla^{2}P^{11}+\frac{3}{4}\partial_{2}\partial_{2}P+\frac{1}{2}\partial_{2}\partial_{2}h_{00}-R_{\hspace{1mm}k}^{1\hspace{1.5mm}1k}-\frac{1}{2}\partial_{i}\partial_{j}P^{ij}+\frac{1}{2}R_{ij}^{\hspace{2mm}ij} \right.  \nonumber \\
        &&\left.-\frac{1}{4m^{2}}\partial_{2}\partial_{2}\partial_{i}\partial_{j}P^{ij}\right)\delta^{2}\left(x-y\right), \nonumber \\
        w_{5}&=&\left(\partial^{1}\partial_{i}P^{i2}+\partial^{2}\partial_{i}P^{i1}-\nabla^{2}P^{12}-\frac{3}{2}\partial_{1}\partial_{2}P-\partial_{1}\partial_{2}h_{00}-2R_{\hspace{1mm}k}^{1\hspace{1.5mm}2k}+\frac{1}{2m^{2}}\partial_{1}\partial_{2}\partial_{i}\partial_{j}P^{ij}\right)\delta^{2}\left(x-y\right), \nonumber \\
        w_{6}&=&\left(\partial^{2}\partial_{i}P^{i2}-\frac{1}{2}\nabla^{2}P^{22}+\frac{3}{4}\partial_{1}\partial_{1}P+\frac{1}{2}\partial_{1}\partial_{1}h_{00}-R_{\hspace{1mm}k}^{2\hspace{1.5mm}2k}-\frac{1}{2}\partial_{i}\partial_{j}P^{ij}+\frac{1}{2}R_{ij}^{\hspace{2mm}ij} \right.\nonumber \\ &&- \left. \frac{1}{4m^{2}}\partial_{1}\partial_{1}\partial_{i}\partial_{j}P^{ij}\right) \delta^{2}\left(x-y\right), \nonumber \\
        w_{7}&=&\frac{1}{4}\partial_{i}\partial_{j}P^{ij}\delta^{2}\left(x-y\right), \nonumber \\
        w_{8}&=&\left(\frac{1}{m^{2}}\left(\nabla^{2}K-\partial_{i}\partial_{j}K^{ij}\right)-K-\pi_{i}^{i}\right)\delta^{2}\left(x-y\right),
\end{eqnarray}
hence,  by using   the second-class constraints (\ref{seconclass}) and the Lagrange multipliers (\ref{multiplier}) into   the extended Hamiltonian, it  will take the following form
\begin{eqnarray}
      \mathcal{H}_{E}&=&\frac{m^{2}}{4}P^{ij}P_{ij}-P^{ij}R_{ilj}^{\hspace{3mm}l}+\frac{3}{4}PR_{lm}^{\hspace{3mm}lm}+\frac{2}{m^{2}}\partial^{l}K_{il}\partial^{j}K^{i}_{j}-\frac{4}{m^{2}}\partial^{j}K_{ij}\partial^{i}K+\frac{2}{m^{2}}\partial_{i}K\partial^{i}K \nonumber \\ &&+
       K^{2}-K_{ij}K^{ij}+\frac{1}{2}h^{ij}\left(R_{ikj}^{\hspace{3mm}k}-\frac{1}{2}\delta_{ij}R_{lm}^{\hspace{2.5mm}lm}\right)+2\pi^{ij}K_{ij}+2\pi^{ij}\partial_{i}h_{0j} \nonumber \\
       &&-\left(\frac{1}{m^{2}}\left(\nabla^{2}K-\partial_{i}\partial_{j}K^{ij}\right)-K-\pi_{i}^{i}\right)\left(K+\frac{1}{m^{2}}\left(\partial_{i}\partial_{j}K^{ij}-\nabla^{2}K\right)+\pi_{i}^{i}\right) \nonumber \\
       &&-\frac{1}{2}h_{00}\left(\partial_{i}\partial_{j}P^{ij}-R_{ij}^{\hspace{2.5mm}ij}+\left(\partial_{i}\partial_{j}-\delta_{ij}\nabla^{2}\right)\tau^{ij}\right)-\left(R_{\hspace{1mm}k}^{l\hspace{1.5mm}mk}-\frac{1}{2}\delta^{lm}R_{ij}^{\hspace{2mm}ij}\right)\tau_{lm} \nonumber  
       \\& &+\frac{1}{4}\partial_{i}\partial_{j}P^{ij}\left(P+\frac{1}{m^{2}}R_{lm}^{\hspace{2mm}lm}+\frac{1}{m^{2}}\left(\partial_{l}\partial_{m}-\delta_{lm}\nabla^{2}\right)\tau^{lm}\right)\nonumber \\
       &&+\left(\partial^{i}\partial_{l}P^{lj}-\frac{1}{2}\partial_{l}\partial_{m}P^{lm}\delta^{ij}-\frac{1}{2}\nabla^{2}P^{ij}-\frac{3}{4}P\left(\partial^{i}\partial^{j}-\delta^{ij}\nabla^{2}\right)\right)\tau_{ij},
\end{eqnarray}
we can observe a squared  term in the momenta $\pi$ which implies that  the Hamiltonian is  bounded from below. This means that the two degrees of freedom are physical rather than  ghosts, in agreement with the Lagrangian analysis reported in \cite{4}.  Furthermore, we can see that the inclusion of the second-class constraints fixes the Ostrogradski instability, hence,  for obtaining  detailed   results in the canonical formalism of higher-order theories,  it is mandatory to develop an  analysis as has been done in this work. In this respect, we have added the appendix A where the system  Klein-Gordon-Einstein  (see \ref{eqn:lin}) without the constraint $ \stackbin[Lin]{}{R}=0$ has been analyzed. This theory is described by a higher-order Lagrangian and  we will observe that its  second-class constraints will have a  trivial structure and the Ostrogradski instability will not be removed. \\
We complete our canonical analysis with the Dirac algebra between the first-class constraints and the extended Hamiltonian, this is  given by
\begin{eqnarray}
\nonumber
    \left\lbrace \Gamma_{1}, \mathcal{H}_{E}\right\rbrace_{D}&=&\frac{1}{2}\Gamma_{4},\nonumber \\
    \left\lbrace \Gamma_{2}^{i}, \mathcal{H}_{E}\right\rbrace_{D}&=&\Gamma_{3}^{i}, \nonumber\\
     \left\lbrace \Gamma_{3}^{i}, \mathcal{H}_{E}\right\rbrace_{D}&=&0, \nonumber\\
     \left\lbrace \Gamma_{4}, \mathcal{H}_{E}\right\rbrace_{D}&=&0,
\end{eqnarray}
where we observe that the algebra is closed and the extended Hamiltonian is of first-class as expected. Furthermore, by using the Dirac brackets and the extended Hamiltonian it is possible to find the  the following equations of motion 
\begin{equation}
    \dot{h}_{ij}=\left\lbrace h_{ij}, \mathcal{H}_E\right\rbrace_{D}=2K_{ij}+\partial_{i}h_{0j}+\partial_{j}h_{0i},
\end{equation}
which is the relation between $K_{ij}$ and $h_{ij}$ given in (\ref{extrinsic}), and 
\begin{equation}
    \dot{K}=\left\lbrace K, \mathcal{H}_E\right\rbrace_{D}=-\frac{1}{2}\nabla^{2}h_{00}-\frac{1}{2}R_{ij}^{\hspace{2mm}ij},
\end{equation}
namely
\begin{equation}
    2\dot{K}+\nabla^{2}h_{00}+R_{ij}^{\hspace{2mm}ij}=0, 
\end{equation}
that corresponds to the equation of motion $\stackbin[Lin]{}{R}=0$. Hence, our analysis is complete and extends those results reported in literature. 
\section{Conclusions}
In this paper a detailed canonical analysis of $NMG$ has been performed.  We observed that the introduction of the $K$'s-extrinsic curvature type variables allowed  us to develop the analysis in a more economical way in comparison with the standard $OD$ scheme. The Lagrangian was written as a function of its velocities and the null vectors of the theory allowed us to identify the complete structure of the constraints. The extended Hamiltonian was constructed, then we observed that  the Dirac brackets and second-class constraints helped to identify  if the Ostrogradski instabilities were present or not. We observed that the extended Hamiltonian does not present instabilities  and   the theory  describes the propagation of two massive physical degrees of freedom. With all  constraints under control, the results of this work will be useful for performing any progress in the quantization program. In fact, it is well known that the better dynamical description of any system is through the Hamiltonian analysis, then for developing the quantization study it is mandatory to perform a complete classical analysis  as in this work has been done. In this respect, our results also could be extended for studying holography information by using the Dirac brackets. In fact, by coupling $NMG$  minimally  to matter we could to study  if the extended theory presents   non-local Dirac brackets just as massive gravity retains \cite{15}.  \\
On the other hand,  the analysis of $EKG$ and the higher-order term of (\ref{eqn:ac})  were reported  in appendix $A$ and $B$ respectively.  In these theories the second-class constraints had a trivial structure, therefore,  the Poisson brackets and  Dirac ones coincide. In this manner,  the extended Hamiltonians of these theories were not healed from  the Ostrogradski sickness.

\section{Appendix A}
In this appendix we will perform the canonical study of the system whose equations of motion are given by 
\begin{equation}\label{EKGL}
     (\Box +m^2)\stackbin[Lin]{}{G_{\mu\nu}} =0.
\end{equation}
For this system there are not extra constraints. It is straightforward to see that these equations can be obtained from the action 
\begin{equation}\label{KGE}
    S[g_{\mu\nu}]=\int d^{3}x\sqrt{-g}\left(R+\frac{1}{m^{2}}Z\right),
\end{equation}
where 
\begin{equation}
    Z=R_{\mu\nu}R^{\mu\nu}-\frac{1}{2}R^{2}.
\end{equation}
In fact, the equations of motion that emerge from the varion of the  action (\ref{KGE}) are 
\begin{equation}\label{eEKG}
    \frac{3}{2}g_{\mu\nu}R_{\alpha\beta}R^{\alpha\beta}+2RR_{\mu\nu}-\frac{3}{4}g_{\mu\nu}R^{2}-4R_{\mu}^{\alpha}R_{\nu\alpha}+\left(\Box+m^{2}\right)G_{\mu\nu}=0, 
\end{equation}
hence, by considering the  perturbation of the metric around the Minkowski background   into (\ref{eEKG})   the  equations (\ref{EKGL}) are obtained.\\
Furthermore,  if we perform the linearization of the action (\ref{KGE}) we can identify the following linearized Lagrangian 
\begin{equation}\label{KGL}
   \begin{split}
      \mathcal{L}&= \frac{1}{2}\partial_{\alpha} \partial^{\alpha} h_{\mu\nu}\partial_{\alpha} \partial^{\alpha} h^{\mu\nu}+\partial_{\mu}\partial_{\nu}h\partial_{\alpha} \partial^{\alpha} h^{\mu\nu}-\partial_{\mu}\partial_{\alpha}h^{\alpha}_{\nu}\partial_{\alpha} \partial^{\alpha} h^{\mu\nu}-\frac{1}{2}\partial_{\alpha} \partial^{\alpha} h\partial_{\alpha} \partial^{\alpha} h
      \\&\hspace{4mm}+m^{2}\left(\frac{1}{2}\partial_{\mu}h\partial^{\mu}h +\partial_{\mu}h^{\mu\rho}\partial_{\alpha}h^{\alpha}_{\rho}-\frac{1}{2}\partial_{\lambda}h^{\rho\mu}\partial^{\lambda}h_{\mu\rho}-\partial_{\mu}h^{\mu\rho}\partial_{\rho}h\right).
   \end{split}
\end{equation}
We observe that the system corresponds to a higher-order theory as expected. Moreover, if we perform the 2+1 decomposition and the variables (\ref{extrinsic}) are introduced, then the Lagrangian is rewritten in the following new fashion
\begin{eqnarray}
\nonumber 
     \mathcal{L}&=&\dot{K}^{2}-\dot{K}_{ij}\dot{K}^{ij}+m^{2}\left(K^{2}-K_{ij}K^{ij}\right)-2\partial_{i}K\partial^{i}K-2\partial_{i}K^{ij}\partial_{k}K_{j}^{k}+2\partial_{k}K_{ij}\partial^{k}K^{ij} \nonumber \\
     &&+2\partial_{l}K\partial_{m}K^{lm}-\partial_{i}\partial_{j}h_{00}\dot{K}^{ij}+\nabla^{2}h_{00}\dot{K}+\frac{1}{2}\left(R_{ikj}^{\hspace{3mm}k}-\frac{1}{2}\delta_{ij}R_{lm}^{\hspace{2mm}lm}\right)\left(\nabla^{2}+m^{2}\right)h^{ij}
   \nonumber  \\
    &&+\frac{1}{2}R_{ij}^{\hspace{2mm}ij}\left(\nabla^{2}+m^{2}\right)h_{00}+\alpha^{ij}\left(\dot{h}_{ij}-2\partial_{i}h_{0j}-2K_{ij}\right).
\end{eqnarray}
The canonical variables of the system are given by  $h_{\mu\nu}$, $\alpha_{ij}$ and $K_{ij}$, and  its corresponding canonical momenta  $\pi^{\mu\nu}$, $\tau^{ij}$ and $P^{ij}$. Hence,  by  performing  the canonical analysis, the main  results are the following: the canonical Hamiltonian is given by 
\begin{equation}\label{HKG}
    \begin{split}
        \mathcal{H}_c&=\frac{1}{4}P^{2}-\frac{1}{4}P_{ij}P^{ij}-\frac{1}{2}\partial_{i}\partial_{j}h_{00}P^{ij}-m^{2}\left(K^{2}-K_{ij}K^{ij}\right)+2\partial_{i}K\partial^{i}K+2\partial_{i}K^{ij}\partial_{k}K_{j}^{k}\\&\hspace{4mm}-2\partial_{k}K_{ij}\partial^{k}K^{ij}-2\partial_{l}K\partial_{m}K^{lm}+2\pi^{ij}K_{ij}-2\partial_{i}\pi^{ij}h_{0j}-\frac{1}{2}R_{ij}^{\hspace{2mm}ij}\left(\nabla^{2}+m^{2}\right)h_{00} \\&\hspace{4mm}-\frac{1}{2}\left(R_{ikj}^{\hspace{3mm}k}-\frac{1}{2}\delta_{ij}R_{lm}^{\hspace{2mm}lm}\right)\left(\nabla^{2}+m^{2}\right)h^{ij}, 
    \end{split}
\end{equation}
where we can observe that there is only a linear term in the canonical momenta $\pi^{ij}$ and this fact will be associated to the presence of ghosts degrees of freedom. \\
On the other hand, the complete set of constraints is given by the following six first-class constraints 
\begin{eqnarray}
  \nonumber 
      \Gamma_{1}&:&\pi^{00}\approx0, \nonumber\\
      \Gamma_{2}^{i}&:&\pi^{0i}\approx0, \nonumber\\ 
      \Gamma_{3}^{i}&:& \partial_{j}\pi^{ij}\approx 0,  \nonumber
    \\
    \Gamma_{4}&:&\partial_{i}\partial_{j}P^{ij}+\left(\nabla^{2}+m^{2}\right)R_{ij}^{\hspace{2mm}ij} -\left(\nabla^{2}+m^{2}\right)\left(\partial^{i}\partial^{j}-\delta^{ij}\nabla^{2}\right)\tau^{ij}\approx0,
\end{eqnarray}
and the following six second-class constraints  
\begin{eqnarray}
      \chi_{1}^{ij}&:&\pi^{ij}-\alpha^{ij}\approx0, \nonumber \\
      \chi_{2}^{ij}&:&\tau^{ij}\approx0.
\end{eqnarray}
We can observe that the second-class constraints have a trivial structure, and it  is easy to observe that the Dirac and   Poisson brackets coincide to each other,  thus we expect that the instability of the Hamiltonian (\ref{HKG}) will be present. In fact, we can use the Dirac brackets and the second-class constraints for calculating  the extended Hamiltonian, we obtain 
\begin{eqnarray}
\nonumber
        \mathcal{H}_{E}&=&\frac{1}{4}P^{2}-\frac{1}{4}P_{ij}P^{ij}-m^{2}\left(K^{2}-K_{ij}K^{ij}\right)+2\partial_{i}K\partial^{i}K+2\partial_{i}K^{ij}\partial_{k}K_{j}^{k} \nonumber \\ 
        &-&2\partial_{k}K_{ij}\partial^{k}K^{ij}-2\partial_{l}K\partial_{m}K^{lm}+2\pi^{ij}K_{ij}+2\pi^{ij}\partial_{i}h_{0j}  \nonumber \\
        &-&\frac{1}{2}\left(R_{ikj}^{\hspace{3mm}k}-\frac{1}{2}\delta_{ij}R_{lm}^{\hspace{2mm}lm}\right)\left(\nabla^{2}+m^{2}\right)h^{ij}+\left(R^{i\hspace{2mm}jk}_{\hspace{1mm}k}-\frac{1}{2}\delta^{ij}R_{lm}^{\hspace{2mm}lm}\right)\left(\nabla^{2}+m^{2}\right)\tau_{ij} \nonumber \\
        &-&\frac{1}{2}h_{00}\left(\partial_{i}\partial_{j}P^{ij}+\left(\nabla^{2}+m^{2}\right)R_{ij}^{\hspace{2mm}ij} -\left(\nabla^{2}+m^{2}\right)\left(\partial^{i}\partial^{j}-\delta^{ij}\nabla^{2}\right)\tau^{ij}\right).
\end{eqnarray}
In this case, we can see that second class constraints do not heal  the instability and there will be ghosts. In fact, the counting of physical degrees of freedom is performed as follows: there are  24 canonical variables, six first-class and six second-class constraints, so
\begin{equation*}
    DOF=\frac{1}{2}\left(24-6-2*6\right)=3,
\end{equation*}
now there are two modes of helicity $\pm2$ and one mode of zero  helicity being a ghost. Our results complete and extend those reported in \cite{4}.
\section{Appendix B}
In this appendix we will resume the canonical analysis of the higher-order term given in the action (\ref{eqn:ac}), this term is expressed  by 
\begin{equation}
\label{eqn:ac2}
 S[g_{\mu\nu}]=\int d^{3}x\hspace{0.1cm}\sqrt{-g}\left( R_{\mu\nu}R^{\mu\nu}-\frac{3}{8}R^{2} \right).
\end{equation}
By performing  the linearization and the change of variables given in (\ref{extrinsic}),  we find the following Lagrangian 
\begin{eqnarray}\label{lagrangianJ}
         \mathcal{L}_{J}&=&\dot{K}^{ij}\dot{K}_{ij}-\frac{1}{2}\dot{K}^{2}-\frac{1}{2}\dot{K}\nabla^{2} h_{00}+\dot{K}^{ij}\partial_{i}\partial_{j}h_{00}+2\dot{K}_{ij}R_{\hspace{1mm}l}^{i\hspace{2mm}jl}-\frac{3}{2}\Dot{K}R_{ij}^{\hspace{2mm}ij}-2\partial^{l}K_{il}\partial^{j}K^{i}_{j}+4\partial^{j}K_{ij}\partial^{i}K\nonumber \\ 
         &-&2\partial_{i}K\partial^{i}K+R_{ikj}^{\hspace{4mm}k}R_{\hspace{1mm}l}^{i\hspace{2mm}jl}+\partial_{i}\partial_{j}h_{00}R_{\hspace{1mm}k}^{i\hspace{2mm}jk}-\frac{3}{8}R_{ij}^{\hspace{2mm}ij}R_{lm}^{\hspace{3mm}lm}+\frac{1}{8}(\nabla^{2} h_{00})^{2}-\frac{3}{4}\nabla^{2} h_{00}R_{ij}^{\hspace{2mm}ij}+\alpha^{ij}(\dot{h}_{ij}\nonumber \\
         &-&2\partial_{i}h_{0j}-2K_{ij}).
 \end{eqnarray}
Hence, the  results from the canonical analysis are the following: the canonical Hamiltonian is given by 
\begin{eqnarray}
\nonumber 
       \mathcal{H}_{J}&=&\frac{1}{4}P^{ij}P_{ij}+\frac{1}{4}P\nabla^{2} h_{00}-\frac{1}{2}P^{ij}\partial_{i}\partial_{j}h_{00}-P^{ij}R_{ilj}^{\hspace{3mm}l}+\frac{3}{4}PR_{lm}^{\hspace{3mm}lm}+2\partial^{l}K_{il}\partial^{j}K^{i}_{j}\nonumber 
       \\&-&4\partial^{j}K_{ij}\partial^{i}K+2\partial_{i}K\partial^{i}K+\frac{1}{4}\nabla^{2} h_{00}R_{ij}^{\hspace{2mm}ij}+2\pi^{ij}K_{ij}-2 \partial_{i}\pi^{ij}h_{0j}.
\end{eqnarray}
where we observe again the linear term in the $\pi^{ij}$ momenta. Furthermore, the complete set of the constraints is given by the following eight first-class constraints 
\begin{eqnarray}
\nonumber 
      \Gamma_{1}&:&\pi^{00}\approx0, \nonumber \\
      \Gamma_{2}^{i}&:& \pi^{0i}\approx0, \nonumber 
      \\
      \Gamma_{3}^{i}&:&\partial_{j}\pi^{ij}\approx 0, \nonumber \\
     \Gamma_{4}&:& \partial_{i}\partial_{j}P^{ij}\approx 0,  \nonumber \\
     \Gamma_{5}&:&\partial_{i}\partial_{j}K^{ij}-\nabla^{2}K+\pi_{i}^{i}\approx 0,\nonumber 
    \\
    \Gamma_{6}&:&P+R_{lm}^{\hspace{2mm}lm} -\left(\partial_{i}\partial_{j}-\delta_{ij}\nabla^{2}\right)\tau^{ij}\approx0, 
\end{eqnarray}
and the following six  second-class constraints  
\begin{eqnarray}
\nonumber       \chi_{1}^{ij}&:&\pi^{ij}-\alpha^{ij}\approx0, \nonumber \\
      \chi_{2}^{ij}&:&\tau^{ij}\approx0.
\end{eqnarray}
It is worth commenting, that  the second-class constraints have a trivial form just like the system analyzed in appendix A, therefore,  the Dirac brackets will be trivial. In this manner, with the results obtained in previous sections we expect that the system (\ref{eqn:ac2}) will present the Ostrogradski sickness. In this respect, we can carry out the counting of physical degrees of freedom as follows: there are 24 canonical variables, eight first-class constraints and six second-class constraints, thus  
\begin{equation*}
    DOF=\frac{1}{2}\left(24-6-2*8\right)=1, 
\end{equation*}
this degree of freedom corresponds to a ghost. \\
We would like to thank Daniel Grumiller for  the comments and cites recommended.




\end{document}